# Denoising and feature extraction in photoemission spectra with variational auto-encoder neural networks

**Francisco Restrepo**[1*], **Junjing Zhao**[1], **Utpal Chatterjee**[1*]

[1]Department of Physics, University of Virginia, Charlottesville, Virginia 22904, USA
*Corresponding authors: eka4va@virginia.edu (F. R.), uc5j@virginia.edu (U. C.)

**Abstract**

In recent years, distinct machine learning (ML) models have been separately used for feature extraction and noise reduction from energy-momentum dispersion intensity maps obtained from raw angle-resolved photoemission spectroscopy (ARPES) data. In this work, we employ a shallow variational auto-encoder (VAE) neural network to demonstrate the prospect of using ML for both denoising of as well as feature extraction from ARPES dispersion maps.

**Introduction**

In the conventional ARPES setup, photoemitted electrons enter a spherical deflector analyzer—where they are sorted out according to their emission angle and kinetic energy—and subsequently impinge on a two-dimensional detector, which records their energy-momentum distribution[1]. Thus, the detector effectively yields intensity maps from which the dispersion of single-particle states is immediately apparent. This band-mapping capability makes ARPES one of the most powerful techniques in the study of solid-state systems. However, such intensity maps are inevitably plagued by statistical noise, which can be reduced at the cost of long data acquisition times and possible sample degradation in poor vacuum conditions. Moreover, the combined effect of spectrally broad light sources and analyzer optics tends to lower the energy resolution of the measurements. These two effects may obscure features in the intrinsic electronic structure, such as energy gaps, band crossings, many-body renormalizations, or the curvature of weakly dispersing bands.

In experiments involving electron or photon detection, denoising is typically achieved by means of smoothing algorithms like moving average, Loess, Savitsky-Golay, among others. These methods indeed reduce pixel-to-pixel fluctuations, but also tend to introduce long-range variations absent in the original spectrum and, in some cases, broaden or shift spectral features. Resolution broadening is a more challenging problem, as the mathematical operation of extracting the intrinsic spectrum from a convolution integral is highly non-trivial. Although maximum-entropy-based methods like Lucy-Richardson deconvolution[2] and the iterative deconstruction algorithm[3] have enjoyed some success in the analysis of ARPES data, these methods depend on the form of the instrument resolution function, which is not known *a priori*. Other spectral sharpening algorithms include minimum gradient[4] and maximum curvature[5] analyses, which are highly sensitive to noise and may yield distribution curves which are negative in the vicinity of the sharpened features.

Recently, ML techniques have emerged as promising tools in image processing and the analysis of spectroscopic data. For example, Peng et al.[6] considered a shallow convolutional neural network (CNN)—originally proposed by Dong et al.[7]—in the problem of band tracing (or feature extraction) in ARPES. In this work, the authors demonstrated how their network could sharpen bands to such an extent that features which were hidden in the raw data could be revealed. The authors further showed that their CNN could outperform common band tracing methods like maximum curvature and minimum gradient, even in the presence of resolution broadening. On the other hand, Kim et al.[8] showed that a deep residual neural network (RESNET)[9] could minimize statistical noise from ARPES band maps. In contrast to the network of Peng et al., their RESNET was not designed to locate and sharpen peaks in dispersion maps, but to denoise these maps while keeping the underlying intensity profiles intact.

Here, we draw upon the lessons learned from the above works to propose a convolutional VAE network that can simultaneously perform denoising and feature extraction from ARPES dispersion maps. We focus on data from two compounds: the transition metal dichalcogenide 1$T$-TiSe$_2$ and the high-temperature superconductor Bi$_2$Sr$_2$CaCu$_2$O$_{8+\delta}$ (Bi2212). Rather than training a single model to process data from both samples, we have trained the same network with different data sets to perform different tasks. For Bi2212, we train the network to remove noise only. The goal is to demonstrate that noise removal by itself can be instrumental to detect renormalization effects, which are not easily discernible from the raw data. On the other hand, for 1$T$-TiSe$_2$, we train the same network to remove noise and sharpen the peaks, where the goal is to detect subtle features—like peak dispersion and band curvature—in the electronic excitation spectrum.

**Variational auto-encoder network**

A schematic of the neural network employed in this work is depicted in Fig. 1(a). The encoder block consists of three convolutional layers with filters of size $(1, 3, 3, 128)$, $(1, 6, 6, 64)$, and $(1, 15, 15, 32)$, respectively, with a 2-D max pooling before the last operation. The bottleneck of the network is reached with another 2-D max pooling operation. The decoder block consists of three (transpose) convolutional layers with filters of dimensions $(1, 15, 15, 32)$, $(1, 6, 6, 64)$, and $(1, 3, 3, 128)$, where the first two operations are preceded by 2-D up-sampling operations. Finally, a $(1, 1, 1, 1)$ filter is applied to recover the 2-D image. All convolutional layers of the autoencoder are followed by batch normalization and rectified linear unit (ReLU) activation functions, while the last layer is followed by a sigmoid activation function. All convolutions are performed without padding—that is, the "padding" option is always set to "same". We use a minimum square error loss function:

$$L(\boldsymbol{X}, \boldsymbol{F}(\boldsymbol{X}, \boldsymbol{\theta})) = \frac{1}{N^2}\sum_{i,j=1}^{N}(X_{i,j} - F(X, \theta)_{i,j})^2, \tag{1}$$

where $X_{i,j}$ and $F(X, \theta)_{i,j}$ denote pixels in the label and output images, respectively, and $\boldsymbol{\theta}$ represents the total set of weights and biases in the network.

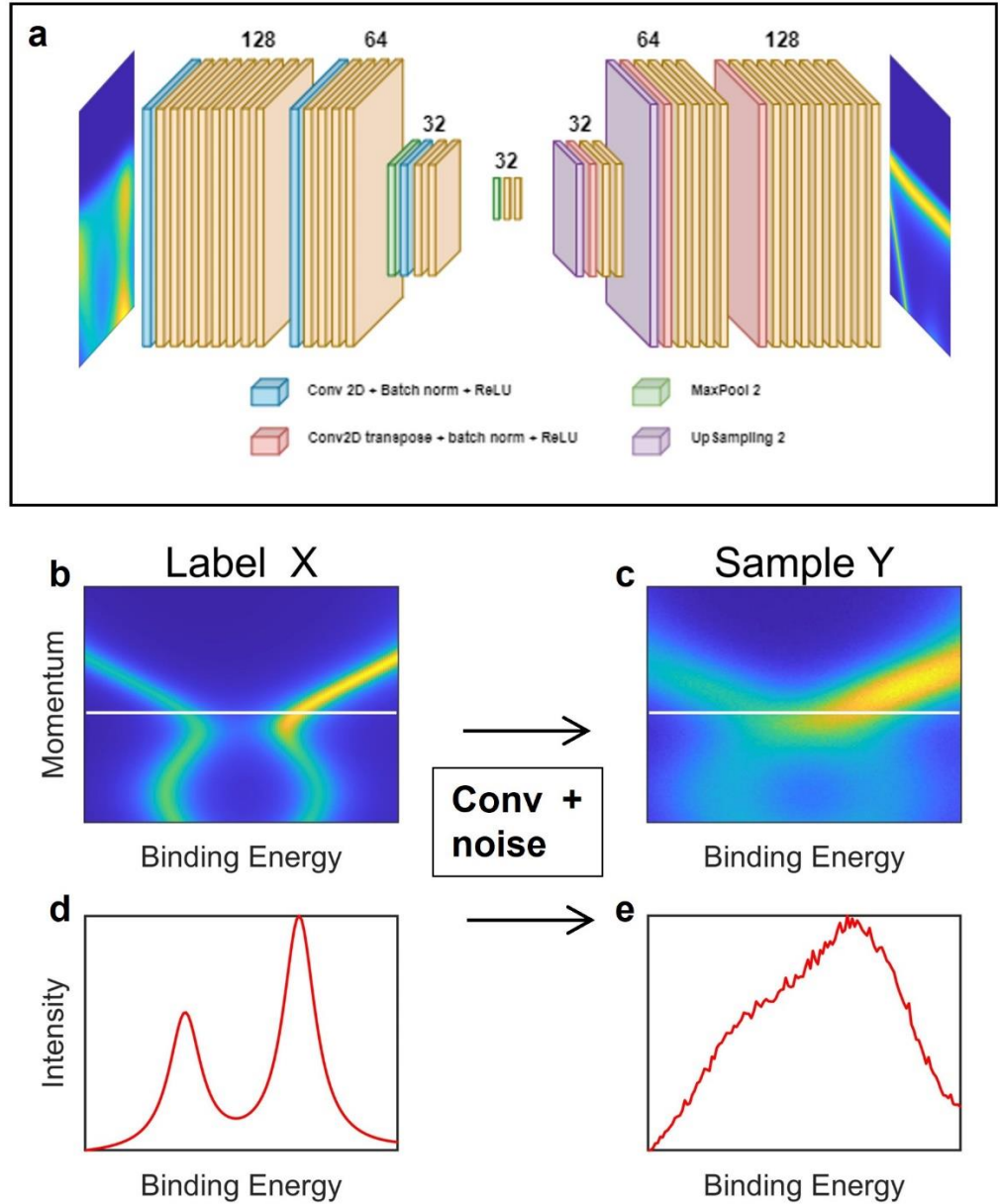

**Fig. 1**: Training the VAE network. (a) Schematic of the proposed VAE network architecture. The operations are color coded as indicated and the numbers on the top indicate the array depth after each operation. (b) The label image **X** is a 128 × 128 patch cropped from a synthetic dispersion representative of 1$T$-TiSe$_2$ near the $\overline{\text{M}}$ point. (c) The sample image **Y** is the result of convolving **X** with a Gaussian function and applying Poisson noise. (d) and (e) Corresponding Energy Distribution Curves (EDCs) along the directions indicated by the white lines in (b) and (c).

Figure 1 also introduces some of the nomenclature that will be used throughout this article. The *label* **X** in Fig. 1(b) is the ideal image, free of distortions from low resolution or signal-to-noise ratio (SNR). The *sample* **Y** in Fig. 1(c) is the result of applying noise to the image and convolving it with a Gaussian resolution function. Note that the bands in the sample are broadened to an extent that the gap near the center of the image, which is clear in the label, is no longer visible. This is more easily appreciated in Fig. 1(d,e), which shows the Energy Distribution Curves (EDCs) plotted in the direction indicated by the white lines in Fig. 1(b,c). The aim of the VAE we propose is to create a function that takes the sample as an input and generates a *prediction* which is as close to the label as possible. Note that Fig. 1 is illustrative of the problem we address in the case of ARPES data from $1T$-TiSe$_2$, where both noise and broadening need to be reduced.

**Network training**

In the first part of this work, we applied the proposed network to the problem of denoising of ARPES dispersions from the cuprate Bi2212, which are typically characterized by severe (intrinsic) broadening and other renormalization effects coming from strong correlations. To this end, we generated synthetic dispersions with many-body renormalizations and large intrinsic broadening. We considered a simple model for the spectral function proposed by Norman et al.[10] which includes the effects of the interaction of fermionic quasiparticles with a collective bosonic mode. In this model, the self-energy is given by

$$\Sigma(\mathbf{k}, \omega) = \frac{\Gamma}{\pi} \ln \left| \frac{\omega - \Omega_k - \Delta_k}{\omega + \Omega_k + \Delta_k} \right| + i\Gamma\Theta(|\omega| - \Omega_k - \Delta_k), \tag{2}$$

where $\Delta_k = \Delta_0 (\cos k_x - \cos k_y)$ is the d-wave superconducting gap and $\Omega_k = 1.3\Delta_k$ is an approximate momentum dependence for the collective mode energy[11]. Introducing this form of the self-energy into the general expression for the spectral function[12]

$$A(\mathbf{k}, \omega) = \frac{1}{\pi} \text{Im} \frac{Z\omega + \epsilon_k}{Z^2(\omega^2 - \Delta_k^2) - \epsilon_k} \tag{3}$$

(where $Z = 1 - \Sigma/\omega$ and $\epsilon_k$ is the dispersion), we generated label spectra exhibiting i) a "kink" in the dispersion along the $\Gamma - Y$ direction, ii) a strong dip in the EDCs from points in the Brillouin zone connected by the wave-vector $\mathbf{Q} = (\pi, \pi)$, and iii) substantial (intrinsic) broadening of the spectra beyond the dip energy $\Delta_0 + \Omega_0$. Next, we convolved the spectral functions with a Gaussian resolution function of width $\sigma_0 \sim 10$ meV. Finally, this "total" spectral function was multiplied by a Fermi function, to reflect the fact that ARPES can only probe the occupied electronic states. The Fermi edge width was approximated by $\Delta\omega_0 \sim \sqrt{(k_B T)^2 + \sigma_0^2}$.

To generate the wide variety of image labels required for proper network training, we randomly varied most parameters entering the spectral function, as summarized in Table I. We varied the six tight binding parameters for the dispersion $\epsilon_k$ ($c_1$ to $c_6$) around the values suggested by Norman[13]. Table I also lists the mean values and variations in $c_0$ (a factor reducing the overall width of the main band), $\Gamma$ (which controls the strength of renormalization effects), $\Delta_0$ (the maximum superconducting gap) and $k_B T$ (the thermal energy scale).

Table I: Summary of parameters used in the generation of label images for the Bi2212 dispersions. All quantities are displayed as a central value plus a maximum deviation. Label images were generated by randomly varying all parameters within the specified limits. Except for the dimensionless factor $c_0$, all entries are in units of meV.

| $c_0$ | $c_1$ | $c_2$ | $c_3$ | $c_4$ | $c_5$ | $c_6$ | $k_B T$ | $\sigma_0$ | $\Gamma$ | $\Delta_0$ |
|---|---|---|---|---|---|---|---|---|---|---|
| 0.5±0.05 | 130.5±20 | -595±50 | 163.6±20 | -51.9±10 | -112±20 | 51±10 | 5 ± 3 | 10±3 | 40±10 | 32±10 |

Next, we generated noisy versions of the labels to use as our training samples. To this end, we first convolved all the EDCs from each label with a second Gaussian resolution function of width $\sigma_1 = 1$ meV. This small value was chosen because, in the case of Bi2212, the network was trained to perform only denoising, not sharpening. (It is important to note that, in generating the samples, we had to start with the labels *before* multiplying by the Fermi function, since the samples had to be multiplied with a different Fermi function of larger width. This consideration also applies to the analysis for $1T$-TiSe$_2$ presented below.) Then, we multiplied the spectra by a resolution-broadened Fermi function of width $\Delta\omega_1 \sim \sqrt{(k_B T)^2 + \sigma_0^2 + \sigma_1^2}$. Making the reasonable assumption that the electron counts recorded in the detector pixels follow the Poisson distribution[6,8], we applied Poisson noise to the sample images as follows. We first rescaled the pixels in these images to assume integer values in the range [0, $n_{max}$] where $n_{max}$ is a typical maximum pixel value measured in ARPES dispersions. By assumption, we can associate a Poisson distribution $P_{ij}$ to the $i,j$ - th pixel with mean value $n_{ij}$, the pixel count. Thus, to apply Poisson noise, we simply generated new pixel values sampled from each pixel's Poisson distribution. Finally, we normalized the images to assume pixel values between 0 and 1, allowing us to use the sigmoid activation function.

Table II: Summary of parameters used in the generation of sample images for the Bi2212 dispersions. All quantities are displayed as a central value plus a maximum deviation.

| $\sigma_1$ (meV) | $n_{max}$ |
|---|---|
| 1 | 2,000 ± 1,000 |

In the second part of this work, we were interested in extracting the Bogoliubov-like dispersion from the ARPES spectra of $1T$-TiSe$_2$ near the $\overline{M}$ point, where broadening and strong matrix element effects render the curvature and back-bending of the bands difficult to resolve. To train the network to resolve these details, we generated a synthetic data set where the samples exhibited the following characteristics: i) a clear Fermi function cutoff, ii) Bogoliubov-like, back-bending bands, iii) substantial intensity anisotropy in momentum space, reflecting the effect of photoemission matrix elements, iv) spectral broadening due to limited energy resolution, and v) Poisson noise characteristic of photoemission data.

Because of our familiarity with the high-temperature superconducting cuprates, we chose to work with a six-parameter tight-binding model for the low energy band in Bi2212, which we adapted to capture the most salient features in the $1T$-TiSe$_2$ dispersion. This main band is split into two branches by introducing a superconducting (in the case of Bi2212) order parameter, which also induces a back-bending of the branches at the Fermi surface. We also introduced an energy offset to locate the energy splitting below the chemical potential, as observed near the $\overline{\text{M}}$ point in 1$T$-TiSe$_2$. In analogy with the simplified, constant self-energy model for the superconducting state, one can write down a spectral function in this case as[12]

$$A(\boldsymbol{k},\omega) = \frac{1}{\pi}\left[\frac{u_k^2\Gamma}{(\omega-E_k+\omega_0)^2+\Gamma^2} + \frac{v_k^2\Gamma}{(\omega+E_k+\omega_0)^2+\Gamma^2}\right], \qquad (4)$$

where $u_k$ and $v_k$ are the Bogoliubov coherence factors, $E_k = \sqrt{\epsilon_k + \Delta_k}$ is the renormalized dispersion, $\Gamma$ is a constant energy broadening, and $\omega_0$ is the energy offset introduced above. However, we found that the above form is too rigid, in that the intensity ratio of the back-bent portions of the branches to the central portions is too small and does not exhibit much variation for different dispersion cuts in momentum space. Therefore, we relaxed this definition by considering a spectral function of the form

$$A(\boldsymbol{k},\omega) = \frac{w_1 u_k^2\Gamma}{(\omega+E_k+\omega_0)^2+\Gamma^2} + \frac{w_2 v_k^2\Gamma}{(\omega+E_k+\omega_0)^2+\Gamma^2} + \frac{w_3 u_k^2\Gamma}{(\omega-E_k+\omega_0)^2+\Gamma^2} + \frac{w_4 v_k^2\Gamma}{(\omega-E_k+\omega_0)^2+\Gamma^2}, \qquad (5)$$

where $w_i$ are weighting factors controlling the relative intensities of the back-bending and central portions of the top and bottom branches. To simulate the excess intensity near the Fermi level, we also included a weakly dispersing, low-energy band of the form

$$\epsilon_{low} = c_7 + \frac{c_7}{2}[\cos k_x + \cos k_y] - \frac{c_8}{2}[\cos 2k_x + \cos 2k_y] - c_8 \cos k_x \cos k_y \qquad (6)$$

and included the corresponding term in the spectral function (Eq. 5), with the same broadening but without the energy offset. Finally, we multiplied all spectra by a Fermi function of width $\omega_0 \sim k_B T$, since the labels in this case were not resolution-broadened. As before, a variety of labels were created by randomly adjusting the various parameters in the spectral function around their mean values within the bounds indicated in Table III.

Table III: Summary of parameters used in the generation of label images for the 1$T$-TiSe$_2$ dispersions. All quantities are displayed as a central value plus a maximum deviation. Label images were generated by randomly varying all parameters within the specified limits. Entries without units are dimensionless.

| $c_0$ | $c_7$ (meV) | $c_8$ (meV) | $w_1$ | $w_2$ | $w_3$ | $w_4$ | $\Gamma$ (meV) | $\Delta_0$ (meV) | $k_B T$ (meV) |
|---|---|---|---|---|---|---|---|---|---|
| $0.32 \pm 0.03$ | $10 \pm 5$ | $4 \pm 2$ | $1 \pm 0.5$ | $1 \pm 0.5$ | $1 \pm 0.5$ | $1 \pm 0.5$ | 7 | $50 \pm 30$ | $10 \pm 5$ |

Next, we generated broadened, noisy, and anisotropic versions of the labels to use as our training samples. To this end, we first convolved all the EDCs from each label with a Gaussian resolution function of mean width $\sigma \sim 30$ meV. This artificially large broadening was necessary for the modeled bands to approximately capture the behavior seen in the data. Similar to the case of Bi2212, we then multiplied the spectra by a broader Fermi function of width $\Delta\omega_1 \sim \sqrt{(k_B T)^2 + \sigma^2}$. Finally, we multiplied the labels by a "window function":

$$f_k = f_0 + \frac{1}{e^{(k-k_{max})/\Delta k_1}+1} + 1 - \frac{1}{e^{(k-k_{min})/\Delta k_2}+1}, \tag{7}$$

which is essentially a sum of Fermi-like functions defined in momentum space with widths $\Delta k_1$ and $\Delta k_2$. The signs were chosen such that the dispersion intensity is left unchanged between $k_{min}$ and $k_{max}$, but is heavily suppressed outside this range. This simple window function was chosen to approximately capture the effect of matrix element-induced intensity modulations in ARPES dispersions, which is a difficult effect to model from first principles.

In creating these samples, as was done with the labels, several parameters were allowed to vary in order to generate a diverse sample pool. Table IV lists these parameters, again reported as a central value plus a maximum deviation. Note that, in this case, the labels only exhibit intrinsic broadening (through $\Gamma$) while the samples are also resolution-broadened by ~ 30 meV. This is because for 1$T$-TiSe$_2$ the network was trained to recover the unbroadened labels and therefore to learn how to sharpen spectral peaks.

Table IV: Summary of parameters used in the generation of sample images for the 1$T$-TiSe$_2$ dispersions. All quantities are displayed as a central value plus a maximum deviation

| $\sigma$ (meV) | $\Delta k_1$ (Å$^{-1}$) | $\Delta k_2$ (Å$^{-1}$) | $n_{max}$ | $f_0$ |
|---|---|---|---|---|
| 30 ± 10 | 0.06 ± 0.02 | 0.06 ± 0.02 | 2,000 ± 1,000 | 0.005 ± 0.005 |

For the Bi2212 analysis, of the total 6,000 label/sample pairs generated, 5,000 were used as training data, while the remaining 1,000 were used as validation data. For 1$T$-TiSe$_2$, we created 7,000 label/sample pairs for training and 2,000 pairs for validation. In both cases, all images had dimensions of 400 pixels in the energy axis and 300 pixels in the momentum axis. To reduce the computational cost and further increase the variety in the types of features shown to the network, we randomly picked patches of size 128 × 128 from these initial images. In the momentum axis, the pixels corresponding to $k_{min}$ and $k_{max}$ were randomly chosen to lie farther than 20 pixels from either edge of these patches—that is, between the 20$^{th}$ and the 108$^{th}$ pixels.

**Performance on a synthetic data set**

As a preliminary check of the performance of the network, we tested the VAE on a synthetic *test* set representative of the 1$T$-TiSe$_2$, for which we expected the network to recover the labels from noisy, broadened samples. This test set was generated following the same guidelines as the training and validation sets but was not shown to the network during training. In Fig. 2, we see that the network satisfactorily reconstructs not only a dispersion map that looks very similar to the label [Fig. 2(a-c)], but also yields EDCs in close quantitative agreement with those from the label image [Fig. 2(d-f)]. For a more detailed comparison, Fig. 2(h) shows a subset of label and predicted EDCs (red and green curves in panels d and f) on the same axes, where they can be seen to nearly overlap. Fig 2(g) shows the loss curves for both the training and validation sets.

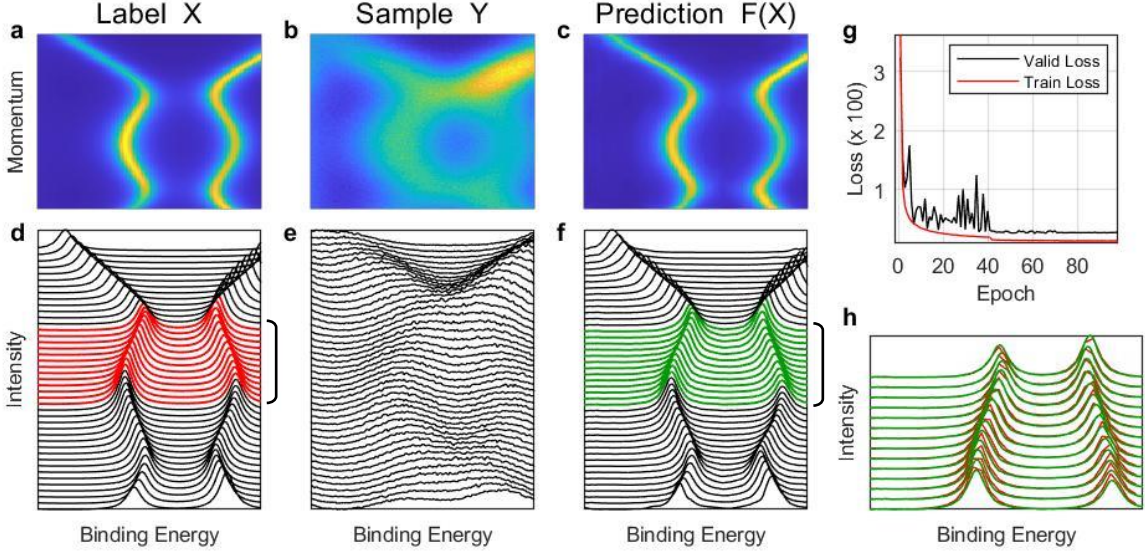

**Fig. 2:** Performance of the VAE network on a synthetic test set not employed during training. (a) – (c) Label, sample, and predicted image for a $128 \times 128$ patch showing the Charge Density Wave (CDW) gap and the backfolding of the branches. Note that the gap is completely removed by resolution broadening in panel (b) and recovered in panel (c). (d) – (f) EDCs corresponding to panels (a) – (c). Red and green curves in (d) and (f) were selected for comparison in panel (h). (g) Training and validation loss curves.

## Performance on raw Bi2212 and 1$T$-TiSe$_2$ dispersion maps

We then tested the performance of the VAE network for the case of ARPES data from Bi2212. Figure 3 shows the results of applying the proposed VAE on two dispersion maps from superconducting Bi2212, corresponding to an underdoped sample ($T_c = 80$ K) collected at different regions in the Brillouin Zone. We begin with a low-count, high-noise dispersion map taken parallel to the $\overline{M} - Y$ direction passing close to the node. Figure 3(a,c) shows the noisy raw data. Figure 3(b,d) corresponds to the output of the VAE. Even though the network was not trained to sharpen features, in Fig. 3b we see that removing the noise enhances the kink feature around 60 meV. Fig. 3(d) shows the same EDCs as Fig. 3(c) after the image was passed through the network. Aside from the perceptual appeal of the smooth curves, note that the dispersion of the top (weaker) band is also easier to track from these EDCs. Figure 3(f-i) shows a similar analysis for a dispersion collected closer to the antinodal region, where strong renormalization effects are present. Following the quasiparticle peaks, we see a sharp dip in intensity, followed by a broad, weakly dispersing hump. The network again produces smoother curves while preserving the intrinsic spectral line shape. This is particularly evident in Fig. 3(e,j), where a selection of sample (red) and predicted (green) EDCs from both dispersions were plotted on the same axes. In both cases, the predicted curves show almost no noise but still represent the underlying line shape quite accurately.

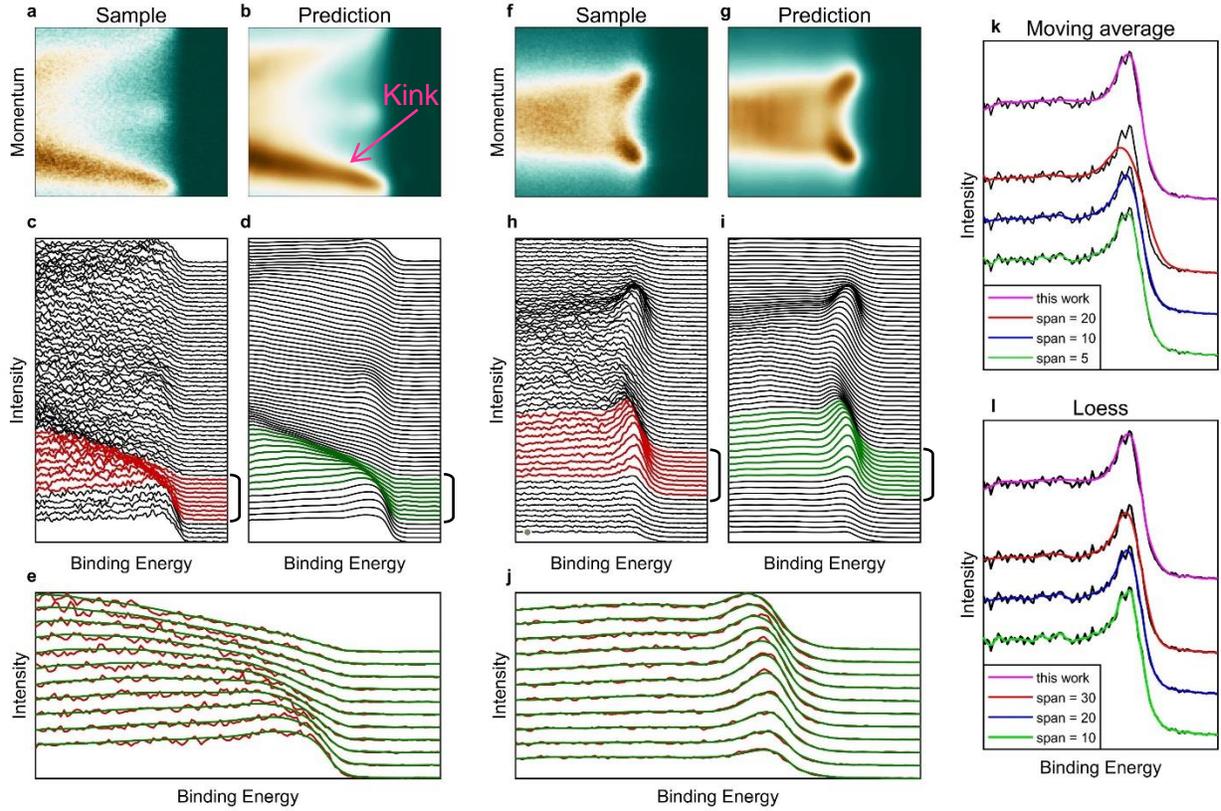

**Fig. 3**: Denoising of Bi2212 spectra (parallel to the $\overline{M} - Y$ direction and below $T_c$) with the proposed VAE network. (a) Dispersion map of slightly underdoped sample ($T_c = 80$ K) near the nodal region. (b) Result of passing image in (a) through the VAE network. (c), (d) EDCs corresponding to (a), (b) respectively. (e) Selected EDCs from (c) (red) and (d) (green) plotted on the same axes. (f) – (j) Same as (a) – (e), but for a cut in a direction parallel to $\overline{M} - Y$ passing closer to the antinode. (k) Comparison with smoothing by "moving average" for three moving window sizes: 5 (green), 10 (blue), and 20 (orange). The pink curve corresponds to the results of the VAE proposed in this work. (l) Same as k, but for the Loess smoothing algorithm. The window sizes are 10 (green), 20 (blue), and 30 (orange).

We also compared the performance of the proposed VAE with two existing smoothing methods: moving average and Loess. Figure 3(k) shows the results for the three window sizes we considered in the moving average smoothing algorithm. For a window size of 5 pixels (green), the algorithm averages the raw data over a small pixel range, so it almost recovers the raw curve with very little smoothing. Increasing the window size to 10 (blue) and 20 (orange) pixels improves the SNR but at the cost of a less accurate representation of the quasiparticle peak. This is particularly evident for the 20-pixel window, where the curve exhibits little noise but the quasiparticle peak looks broader and shifted to higher binding energies. The Loess algorithm performs somewhat better [Fig. 3(l)] but still tends to interpret noisy bumps as intrinsic features, even for the smoothest curve shown (window size 30, red).

We now turn to the transition metal dichalcogenide 1$T$-TiSe$_2$. This compound is known to undergo a Charge Density Wave (CDW) phase transition below the temperature T$_{CDW}$ ~ 200 K. Figure 4(a) shows a schematic representation of the energy bands of 1$T$-TiSe$_2$ along the $\Gamma - \bar{M}$ direction for T > T$_{CDW}$. The maximum of the hole-like valence band derived from Se 4p states is located at the $\Gamma$ point, while the minimum of the electron-like conduction band derived from Ti 3d states is located at the $\bar{M}$ point. In the CDW phase, electron-hole coupling results in the so-called band folding: a replica of the conduction band appears at the $\Gamma$ point, while a replica of the valence band appears at the $\bar{M}$ point [Fig. 4(b)]. A CDW energy gap opens at the points of intersection of the main and replica bands, and the bands bend backward [Fig. 4(c)]. Even though the formation of replica bands has been routinely observed in ARPES data from 1$T$-TiSe$_2$, the back-bending feature is difficult to discern from the data. This can be realized from Fig. 4(d)., where we present data from the Cu-doped TiSe$_2$ sample. Even though the presence of backfolded bands is clear, the back bending feature i.e., the Bogoliubov-like dispersion, is hard to be detected. This is mostly due to resolution-broadening of the spectrum. Therefore, we selected the region around $\bar{M}$ and the chemical potential (magenta box in Fig. 4(d)) for analysis via the proposed VAE [Fig. 4(e,f)]. In Fig. 4(f), the upper and lower branches of the Ti 3d band are much more clearly visible than in the raw data in Fig. 4(e). Moreover, the lower branch exhibits a slight back-bending to higher binding energies which is not obvious from the raw dispersion. This can be more easily appreciated when looking at the EDCs in Fig. 4(g,h). The upper branch, which appears distorted in the raw data due to a strong matrix element effect, is also more clearly visible in the predicted spectra. Note that the predicted EDCs are still positive definite and considerably narrower than the raw ones, so their maxima can be used to track the shape of the band; in particular, the back-bending of the lower branch is immediately apparent in Fig. 4(h). Finally, we note that the outputs in Fig. 4(f,h) also show virtually no noise. Even though a precise line shape analysis is beyond the scope of this work, we note that our VAE yields EDCs with shapes reflecting those used in the training data; namely, Lorentzians. In this context, we would like to point out the results of Peng et al., who trained their network using dispersions consisting only of pixel values 0 and 1, so that their network produced EDCs showing similar discontinuities.

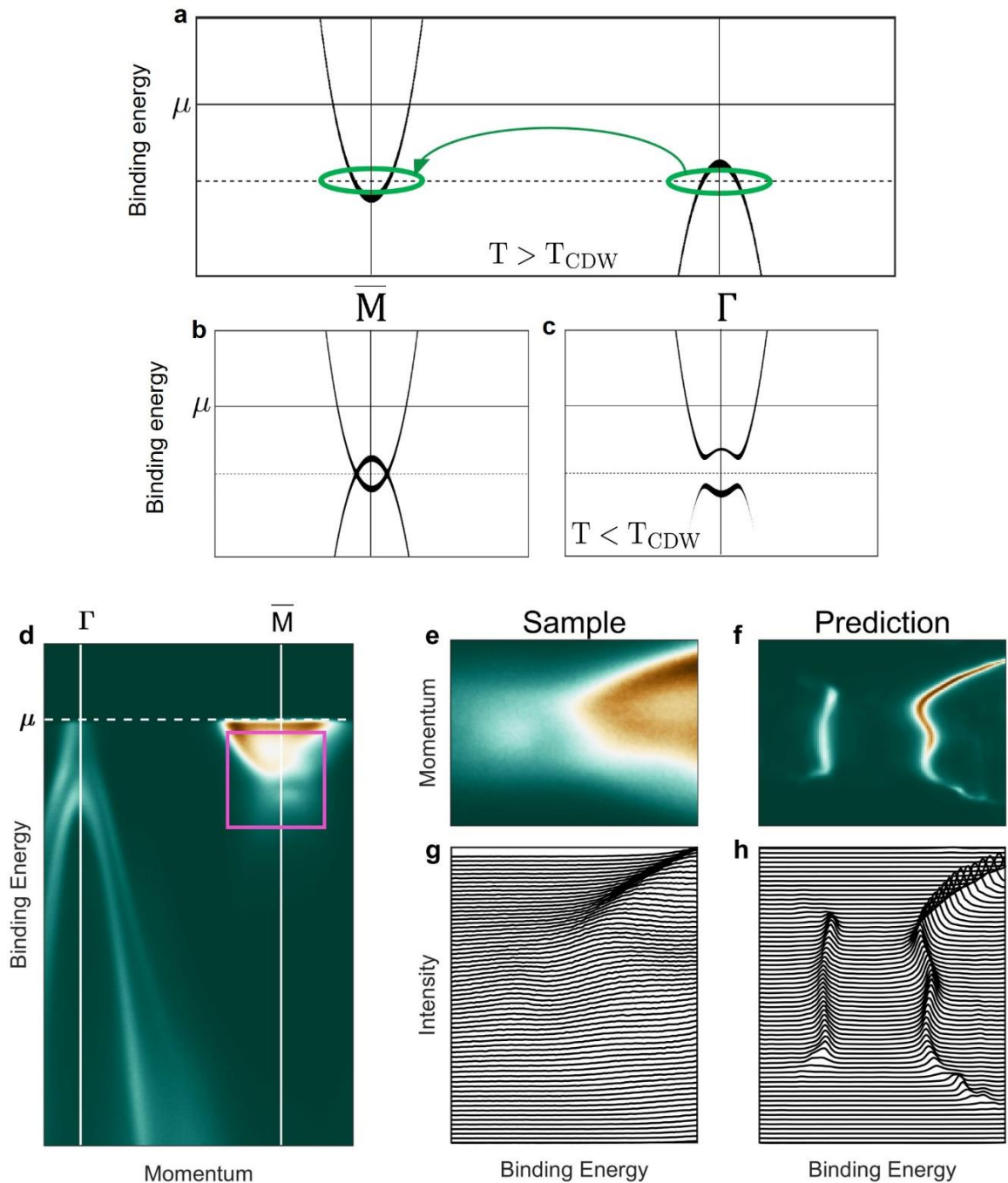

**Fig. 4:** Particle-hole mixing and performance of the VAE network on ARPES data from $1T$-TiSe$_2$. (a) Low energy bands in the $\Gamma - \overline{M}$ direction in the absence of particle-hole mixing (T > T$_{CDW}$). (b) Band crossing resulting from the doubling of the unit cell dimensions induced by particle-hole mixing. (c) Bogoliubov-like dispersion in the CDW phase (T < T$_{CDW}$). (d) Wide energy-momentum dispersion in the $\Gamma - \overline{M}$ direction. The magenta square indicates the patch selected as input to the network. The original and predicted images are shown in panels (e) and (f), respectively (after a clockwise, 90° rotation). (g) – (h) EDCs corresponding to panels (e) and (f), showing the CDW gap and the dispersion more clearly, especially in panel (h).

**Conclusions**

We investigated the performance of a shallow VAE for denoising of ARPES data from Bi2212 and both denoising and sharpening for 1*T*-TiSe$_2$. ARPES data from the cuprate Bi2212 high temperature superconductors were used to demonstrate how the VAE can denoise dispersion maps collected with low SNR values and enhance renormalization features, which are difficult to resolve in the raw data. We believe such denoised dispersions could be the input instead of raw data for other types of analysis, like maximum gradient, minimum curvature, Fourier transform, or even Kramers-Kronig transformation, so as to resolve the complications due to noise in data. To explore the possibility of employing the same network to denoise and sharpen dispersion features, we used data from the transition metal dichalcogenide 1*T*-TiSe$_2$. In this case, we saw that the same network could resolve the dispersion of the top and bottom branches of the Bogoliubov-like bands around the $\bar{\text{M}}$ point of the Brillouin zone. We believe that the capability of the same network to denoise and sharpen dispersion features will be pertinent particularly to ARPES experiments in soft x-ray and hard x-ray regime, where a lack of high resolution and high statistics data stymie reliable extraction of band parameters, making it possible to harness the potential of bulk probe ARPES to its fullest extent.

We further expect that with better theoretical models, more accurate representations of the ARPES data will be available for training, which will yield outputs which more closely match the underlying spectral function. This would be particularly helpful in the study of strongly correlated materials like the cuprate superconductors, where the complex line shapes are governed by an energy-momentum-dependent self-energy.

Our results also highlight the flexibility of the proposed VAE since, by using two different training sets, the network was taught to perform two different tasks. It should be possible to train the VAE to perform similar tasks on data from different compounds if appropriate training data can be generated. This, of course, requires having basic, a priori knowledge of the band structure of the material. Given the short time it took to train the two models discussed in this work (about 20 minutes for denoising and 40 minutes for denoising and sharpening on an RTX – 3090 GPU), we believe this compromise in generality is quite acceptable.

**Acknowledgement**
The authors thankfully acknowledge helpful discussions with Aravind Krishnamoorthy. U.C. acknowledges support from the National Science Foundation under grant number DMREF-1629237.